\documentclass[a4paper,timesnewroman,12pt]{article}
\usepackage[left=2.5cm,right=2.5cm,top=2.5cm,bottom=2.5cm]{geometry}
\usepackage{xcolor}
\usepackage{booktabs} 
\usepackage{graphicx}
\usepackage{amsmath}
\begin{document}
\begin{center}
	{\large\bf 
	A Two-Population Mortality Model to Assess Longevity Basis Risk
	}\\
	
	\vspace{0.5cm}
	Selin \"{O}zen$^1$, \c{S}ule \c{S}ahin$^2$\\
	
	\vspace{0.5cm}
	{\it
	\footnote{selindegirmenci@karabuk.edu.tr} Karabuk University, Department of Actuarial Sciences, Karabuk, TURKEY \\
	\footnote{Sule.Sahin@liverpool.ac.uk}Department of Mathematical Sciences, Institute for Financial and Actuarial Mathematics, University of Liverpool, Liverpool, UK\\

	}
\end{center}
\begin{abstract}
	Index-based hedging solutions are used to transfer the longevity risk to the capital markets. However, mismatches between the liability of the hedger and the hedging instrument cause longevity basis risk. Therefore, an appropriate two-population model to measure and assess the longevity basis risk is required. In this paper, we aim to construct a two-population mortality model to provide an effective hedge against the longevity basis risk. The reference population is modelled by using the Lee-Carter model with renewal process and exponential jumps proposed by Özen and Şahin \cite{ozensahin} and the dynamics of the book population are specified. The analysis based on the UK mortality data indicate that the proposed model for the reference population and the common age effect model for the book population provide a better fit compared to the other models considered in the paper. Different two-population models are used to investigate the impact of the sampling risk on the index-based hedge as well as to analyse the risk reduction regarding hedge effectiveness. The results show that the proposed model provides a significant risk reduction when mortality jumps and the sampling risk are taken into account.
\end{abstract}	
\section{Introduction}
Longevity risk can be defined as the risk that members of some reference population might live on average longer than anticipated. It is a crucial financial concern for both pension plans and life insurers since the institutions might have to make higher payments than expected due to the longevity risk. Life expectancy continues to rise in association with improvements in nutrition, hygiene, medical knowledge, lifestyle, and health care. Uncertainty about future mortality improvements might have significant economic implications for annuity providers, pension providers, and social insurance programs. Although the individuals have different lifespan, longevity risk might affect all pension plans and life insurers, and hence it is not possible to diversify it with an increase in portfolio size. Therefore hedging of the longevity risk is of critical importance for both pension plan providers and life insurance companies

Various solutions have been presented to manage and mitigate the longevity risk. Index-based hedging solutions, which include longevity-linked securities and derivatives, provide more advantages over other hedging solutions, such as faster execution, greater transparency, liquidity potential, and lower costs \cite{li2018}. Due to offering significant capital savings and providing effective risk management, index-based longevity instruments  attract increased interests from within and outside of the worlds of insurance and pensions.

The first step of the assessment of longevity risk and thus the valuation of index-based financial products is the mortality modelling. The choice of the appropriate model is crucial to quantify the risk and provide a foundation for pricing and reserving. Due to the inadequacy of the quality and the size of the portfolio, a reference population index is commonly used by hedgers in index-based hedging solutions. The payments of the financial products are associated with this reference population index, but not the (book) population that underlies the portfolio that is being hedged. Therefore, longevity risk trading usually entails two different populations: the first is affiliated with the portfolio of the hedger, while the other is linked to the hedging instrument \cite{zhou2013}. There would then be a potential mismatch between the hedging instrument and the portfolio, due to certain demographic differences (e.g. age profile, sex, socioeconomic status). This might give rise to longevity basis risk, the assessment of which is under research in the latest actuarial literature \cite{li2018}. Hence, a multi-population mortality model is required to provide an accurate mortality model for measuring the basis risk.

Several multi-population mortality models have recently been presented while only \cite{zhou2013} consider the transitory mortality jump effects in the modelling process. It is important to incorporate the mortality jumps to estimate the uncertainty surrounding a central mortality projection. Incorporating the jumps into the modelling process allows us to estimate  the probability of catastrophic mortality deterioration when pricing securities for hedging extreme mortality risk \cite{zhou2013}. In this paper, a different approach proposed by Özen and Şahin \cite{ozensahin}, has been used for modelling jump effects. This approach includes the history of catastrophic events in the jump frequency modelling process by using renewal process as well as a specification of the Lee--Carter (LC) model for mortality.

The aim of this paper is to build an appropriate two-population mortality model incorporating mortality jumps to assess the longevity basis risk for pricing longevity-linked financial products. Such a model provides a basis to effective risk management strategies. To illustrate the impact of our proposed mortality model in hedge effectiveness, we consider a hedge for a hypothetical pension plan. Moreover, we take sampling risk into account since the available historical data is usually small for a pension plan. Therefore, the size of a pension plan is examined in regard to hedge effectiveness. We also compare the hedge effectiveness of our model with the other two commonly used mortality models. The results show that our proposed model provides a better risk reduction.

The remainder of the paper is structured as follows. Section \ref{sec2} introduces some helpful notations. In Section \ref{sec3}, an overview of the existing multi-population mortality models is provided. The steps for building a two-population mortality model are described in Section \ref{sec4}. Section \ref{sec5} applies the proposed model to a hypothetical pension plan and examines the effectiveness of the hedge. Finally, Section \ref{sec6} concludes the paper.

\section{Notations}\label{sec2}

We begin with introducing some helpful notations adopted from Villegas et al. \cite{villegas2017}. Let us denote the reference population by $R$ that is backing the hedging instrument, and $B$ is used for the book population whose longevity risk is going to be hedged. 
Time will be measured in units of years, and year $t$ will refer to time interval $[t, t+1]$. For the reference population, $D_{x,t}^R$ and $E_{x,t}^R$ show the death counts and exposure to risk at age $x$ at last birthday in year $t$. Central mortality rates for any individual of the reference population of age $x$ in year $t$ will be denoted by $m_{xt}^R$ and computed as $m_{x,t}^R=D_{x,t}^R/E_{x,t}^R$. Likewise, the same values for the book population are given here as $D_{x,t}^B$, $E_{x,t}^B$ and $m_{x,t}^B=D_{x,t}^B/E_{x,t}^B$.

A further assumption being made here is that the data for the reference and book populations can be different regarding specified sets of ages and specified amounts of years. For instance, we have $D_{x,t}^R$ and $E_{x,t}^R$ for consecutive ages $x=x_1,...,x_{n_R}$ and consecutive calender years $t=t_1,...,t_{n_R}$ in the reference population, while $D_{x,t}^B$, $E_{x,t}^B$ are available for ages $x_1,...,x_{n_B}$ and calender years $t=u_1,...,t_{n_B}$ in the book population.

The reference population's data might be provided for a longer time frame than that of the book population, which is $n_R\geq n_B$. Moreover, the calendar years of data in a book may be provided as a subset of the comparable calendar years for the reference population, $t_{n_B} \ne t_{n_R} $. Also, the ages provided by the book might constitute a smaller portion of those that are provided for the reference population.
\section{An Overview of Mortality Models for Measuring Basis Risk}\label{sec3}

We need to specify an appropriate two-population model for $m_{x,t}^R$ and $m_{x,t}^B$ which has the ability to capture the trends present within both the book and reference populations. It is crucial to incorporate these trends since the mortality trends of the reference population support the hedging instrument while the trends in the book population are significant for the longevity basis risk to be hedged. Future mortality will be forecasted by the specified model in a consistent way. 

Several models have been developed to display the mortality evolution of two related populations. These models usually derived by expanding the previous single-population models by incorporating the correlations and interactions existing between populations. Although the majority of research on modelling multi-population has been conducted relatively recently, the seeds are traced back to the influential paper published by Carter and Lee \cite{carter1992}. The paper introduced feasible approaches for the extension of the authors' single-population model for differences in US mortality between men and women. The model suggests applying independent Lee--Carter models to individual populations as the first approach for multi-population modelling. Afterwards, the joint-$\kappa$ model, based on the assumption that populations' mortality dynamics are driven by one commonly shared time-varying factor, was developed. The third approach was based on an extension of the Lee--Carter model, applying co-integration techniques and estimating the populations jointly. Brief descriptions of the new models established on the basis of the Lee--Carter model are given below:\\\medskip

\noindent {\textit{Independent Modelling:}} In this approach, mortality is modelled with the utilisation of two independent Lee--Carter models. Let $m_{x,t}^i$ be the central death rate for population $i$ in year $t$ at age $x$. The model can then be expressed as follows:
\begin{equation}\label{eq1}
\ln (m_{x,t}^i)=a_x^i+b_x^ik_t^i+e_{x,t}^i, \quad i=R,B.
\end{equation}

All of those parameters hold the same meanings that they possess in the original Lee--Carter model. It is possible to estimate the model parameters with the application of singular value decomposition, the Markov Chain Monte Carlo approach, or maximum likelihood estimation. A mortality index can be modelled using two independent ARIMA processes for forecasting purposes. Although the model is easily applicable it ignores the dependency between the mortality rates of the populations. Hence, it might lead an overestimation of the basis risks.\\\medskip

\noindent \textit{The Joint-k Model:} This model is based on the assumption that the mortality rates of both populations being driven by one single mortality index. This model may be expressed in the following way:

\begin{equation}\label{eq2}
\ln (m_{x,t}^i)=a_x^i+b_x^ik_t+e_{x,t}^i, \quad i=R,B.
\end{equation}

In the joint-k model, the mortality index is the driving force of the changes in mortality rates for both populations. Model parameters are estimated as in the previous approach while the mortality index $k_t$ is modelled based on an appropriate ARIMA process. However, the model assumes that the mortality improvements of the populations are perfectly correlated and the existence of the common factor suggests identical advancements in mortality for both populations for all periods. Hence, the assumption is not realistic. \cite{li2005} introduced a population-specific factor for this model, which is referred to as the ``augmented common factor model''.\\\medskip

\noindent \textit{Augmented Common Factor:}  For the first approach, that of the two independent Lee--Carter models, life expectancy divergence increases in the long run. The joint-k model cannot completely resolve this issue, since discrepancy between two populations in terms of parameter $b_x^i$  could generate divergences in the mortality predictions.

Li and Lee \cite{li2005} present criteria for the divergence problem, as given below:
\begin{itemize}
	\item[-] $b_x^R$ = $b_x^B$ for all x.\\
	\item[-] $k_t^R$ and $k_t^B$ have identical drift terms of the ARIMA process.  
\end{itemize}

Given these conditions, Li and Lee \cite{li2005} introduced a specific factor for the Lee--Carter model:

\begin{equation}\label{eq3}
\ln (m_{x,t}^i)=a_x^i+b_xk_t+b_x^ik_t^i+e_{x,t}^i, \quad i=R,B.
\end{equation}

$b_x^ik_t^i$ term serves to capture variations in the changing rate of mortality of population $i$ from the long-term mortality change tendencies suggested by the common factor, $b_xk_t$. The $k_t^i$ factors are modelled using the AR(1) process to ensure the avoidance of any divergence from the mortality projections \cite{li2011}. \\\medskip

Another modelling approach for two-population mortality is the extension of the Cairns--Blake--Dowd (CBD) mortality model for a single population \cite{cairns2006}. A version of the CBD model for two populations and its variants were introduced by Li et al. \cite{lietal2015}. For example the two-population variant of the CBD model with the incorporation of quadratic effects, known as the M7 model, can be described as follows: 

\begin{equation}\label{eq4}
\text{logit}\ q_{x,t}^i=\kappa_t^{i,1}+(x-\bar{x})\kappa_t^{i,2}+\big((x-\bar{x})^2-\sigma_x^2\big)\kappa_t^{i,3}+\gamma_{t-x}^i, \quad i=R,B,
\end{equation}

\noindent where $\bar{x}$ denotes average age and $\sigma_x^2$ is the average value of $(x-\bar{x})^2$. $\kappa_t^{1,i}$ and $\kappa_t^{2,i}$ are two stochastic processes which represent the two time indices of the model. Time index $\kappa_t^{1,i}$ reflects the level of mortality measured at time $t$, while  $\kappa_t^{2,i}$ shows the slope and affects every age differently. $\gamma_{t-x}^i$ parameter represents the cohort effect. Li et al. \cite{lietal2015} considered three different approaches, which were presented in the work of  \cite{zhou2014} to forecast future mortality rates.

The use of an age-period-cohort (APC) model with two populations was presented by Cairns et al. \cite{cairns2011} and Dowd et al. \cite{dowdetal2011}. The model is expressed in the following way:

\begin{equation}\label{eq5}
\log m_{x,t}^i=a_x^i+k_t^i+\gamma_{t-x}^i, \quad i=R,B.
\end{equation} 
$a_x^i, k_t^i$ and $\gamma_{t-x}^i$ are the age, period and cohort effects of the populations.

Spreads that exist between the state variables can be modelled as a mean-reverting process for each population so that the short-term trends in the mortality rates can vary, whereas there are parallel long-term improvements. In Cairns et al. \cite{cairns2011}, a Bayesian framework which allows to estimate non-observable state variables and the underlying parameters of the stochastic process in one stage, is used. Moreover, Dowd et al. \cite{dowdetal2011} developed a gravity approach in which the mortality rates of two populations experience attraction to one another determined by a dynamic gravitational force. The force depends on the comparative sizes of the populations in question \cite{villegas2017}.

Jarner and Kryger \cite{jarner2011} and Cairns et al. \cite{cairns2011} recognised the comparative value of the reference population supporting the index and the population whose longevity risk is being hedged. Their approach centres on the reference population at the beginning, after which the dynamics of book mortality must be given for the incorporation of characteristics from the reference population. This relative method has important aspects such as it permits the mismatching of data between the book and reference population and it is applicable in the typical case in which a book population is significantly smaller than a reference population \cite{haberman2014}. The mortality models that are used in the relative method are presented in Table \ref{Tab1} \cite{villegas2017}.

\begin{table}[h]
	\caption{Mortality Models for The Relative Method}
	\centering
{\scriptsize 	\begin{tabular}{ccccccccccc}
		\toprule
		Original& Model Name & Reference& Book-Reference\\
		Model & & Population & Difference Formula\\
		\midrule
		Common Factor & CF+Cohorts&LC+Cohorts&$a_x^B$\\
		Common Age Effect&CAE+Cohorts&LC+Cohorts&$a_x^B+\beta_x^Rk_t^B$\\
		Relative LC with Cohorts& RelLC+Cohorts&LC+Cohorts&$a_x^B+\beta_x^Bk_t^B$\\
		Gravity&Gravity (APC)& APC & $a_x^B+k_t^B+\gamma_{t-x}^B$\\
		Two-population M5&M7-M5& M7 & $\kappa_t^{(1,B)}+(x-\bar{x})\kappa_t^{(2,B)}$\\
		Two-population M6&M7-M6& M7& $\kappa_t^{(1,B)}+(x-\bar{x})\kappa_t^{(2,B)}+\gamma_{t-x}^B$\\
		Two-population M7&M7-M7&M7&$\kappa_t^{(1,B)}+(x-\bar{x})\kappa_t^{(2,B)}+((x-\bar{x})^2-\hat{\sigma}_x^2)\kappa_t^{(3,B)}+\gamma_{t-x}^B$\\
		Saint Model&M7-Saint&M7&$\kappa_t^{(1,B)}+(x-\bar{x})\kappa_t^{(2,B)}+((x-\bar{x})^2-\hat{\sigma}_x^2)\kappa_t^{(3,B)}$\\
		Plat Relative Model&M7-Plat&M7&$\frac{100-x}{100-\bar{x}}\kappa_t^{(1,B)}$\\	
		\bottomrule
		\label{Tab1}
	\end{tabular}}
\end{table} 

There are other multi-population applications of well-known single-population models. For instance, Biatat and Currie \cite{biatat2010} expanded the P-spline approach to encompass scenarios with two populations; previously, it had been utilised with success for cases of single populations. Hatzopoulos and Haberman \cite{hatzopoulos} and Ahmadi and Li\cite{ahmadi} applied a multivariate generalised linear model (GLM) for obtaining coherent forecasting of mortality in cases of multiple populations \cite{villegas2017}.

However, to our knowledge, only Zhou et al. \cite{zhou2013} incorporates jumps that are due to interruptive events such as the Spanish flu epidemic in 1918 to two-population mortality model. Their model can be regarded as a two-population generalisation of the model in Chen and Cox \cite{chencox2009}. They assumed that the mortality of a population is either jump-free or subject to one transitory mortality jump. The severity of a mortality jump is normally distributed.

Although many multi-population mortality models exist, only a few investigates how to measure the longevity basis risks. Some of the earlier research designed for quantifying basis risk, such as Cairns et al. \cite{cairnsetal2014}, Ngai and Sherris \cite{ngaisherris}, and Li and Hardy \cite{li2011}, have applied the original framework constructed by Coughlan et al. \cite{coughlan2011}.

\section{Building a Two-Population Mortality Model}\label{sec4}
The first step in pricing the longevity-linked products is to establish a two-population mortality model in order to measure the longevity basis risk. A relative approach is applied in this paper, as in Haberman et al. \cite{haberman2014}, since it has many advantages over joint modelling. However, the modelling framework is slightly different from the original formulations used for the reference model.

\subsection{Mortality Data}

All of the examples provided in the paper utilise historical UK mortality data, which were collected from the Continuous Mortality Investigation (CMI) and the Human Mortality Database (HMD). The first data represent the mortality experience of CMI assured male lives that are being hedged. The subsequent dataset is for the reference population, which provides the mortality experience of male lives in England and Wales (EW). For the reference population, a sample period from 1961 to 2016 is considered, while for the book population, the sample period comprises the years of 1961-2005. The sample age range being considered is 65 to 89.

\subsection{Modelling the Reference Population}

The model considered in the paper is a Lee--Carter model with exponential transitory jumps and renewal process. By using renewal process, we attempted to include the history of catastrophic events into the mortality modelling process. In Özen and Şahin \cite{ozensahin}, the proposed model was compared to other mortality models with jump effects. The analysis has shown that the arrivals between two catastrophic events is important and the proposed model provides a better fit to the historical data (see Özen and Şahin \cite{ozensahin} for more details). Moreover, as indicated before, mortality jumps have important impacts on mortality dynamics and it is essential that they are incorporated into the modelling process. Hence we use the Lee--Carter model with exponential transitory jumps and renewal process as our reference population mortality model.

Here, we assume that transitory jumps are only valid for the reference population because of the quality and size of the available data for the national population. The proposed model is given by the following:

\begin{equation}\label{eq6}
log(m_{x,t}^R)=a_x^R+b_x^Rk_t^R,
\end{equation}

\begin{equation}\label{eq7}
k_t^R=k_0^R+(\mu-\frac{1}{2}\sigma^2)t +\sigma W(t)+ \sum_{i=1}^{N(t)}Y_i.
\end{equation}

Here, $m_{x,t}^R$ denotes the central death rate in year $t$ for age $x$, $a_x^R$ represents the age pattern of the death rates, $k_t^R$ reflects variations that exist across time in the log mortality rates, $b_x^R$ represents the mortality rates' sensitivity to changes in time-varying mortality index $k_t^R$, $W(t)$ signifies standard Brownian motion, $N(t)$ denotes the renewal process, and, finally, $Y(i)$ denotes a sequence of iid exponential random variables representing the size of the jumps. 

There are two identifiability constraints, which means that unique solutions exist for all of the model's parameters. These identifiability constraints are given as follows:
$$\sum_{x}b_x^R=1, \qquad \sum_{t}k_t^R=0. $$
We will estimate the model's parameters using the MLE method. First, reference population parameters $a_x^R$, $b_x^R$, and $k_t^R$ are estimated. Afterwards, Equation (\ref{eq7}) is used to calibrate the time-varying mortality index. We need to find the density function of the independent one-period increments, $\Delta k_i^R=r_i=k_i^R-k_{i-1}^R$, to estimate the parameters of the calibrated model.

Let $D=\{k_0,k_1,...,k_T\}$ represent the mortality time series at times of $t=1,2,...,T$, which have equal spacing. The one-period increments are independent and identically distributed (iid). Unconditional density for the one-period increment $f(r)$ may be given as follows:
\begin{equation}\label{eq8}
f(r_i)=P(0)f(r_i|0)+\sum_{n=1}^{N(t)}P(n)f(r_i|n),
\end{equation}
where $P(0)=1-F(t)$, $P(n)=\int_{0}^{t}P_{n-1}(t-s)f(s)ds$ are the probability of no jump and $n$ jumps occur in the renewal process, where $F(t)$ and $f(t)$ are the distribution and density functions of inter-arrival times of between two jumps. $f(r_i|0), f(r_i|n)$ are conditional densities for a one-period increment; more specifically, they are conditional on the given numbers of jumps and expressed as:

$$f(r_i|0)=\frac{1}{\sqrt{2\pi}\sigma}e^{-\frac{(r-\mu+0.5\sigma^2)^2}{2\sigma^2}}$$

$$f(r_i|n)=\int_{0}^{\infty}\frac{\eta^n}{(n-1)!}X^{n-1}e^{-\eta X} \frac{1}{\sqrt{2\pi}\sigma} e^{-\frac{(r-X-\mu+0.5\sigma^2)^2}{2\sigma^2}}dx$$
$$=\frac{\eta^n}{(n-1)!\sqrt{2\pi}\sigma}\int_{0}^{\infty}X^{n-1}e^{-\eta X-\frac{1}{2\sigma^2}(r-X-\mu+0.5\sigma^2)^2}dx$$

\begin{figure}[h]
	
	\centering
	\includegraphics[width=9cm, scale=1]{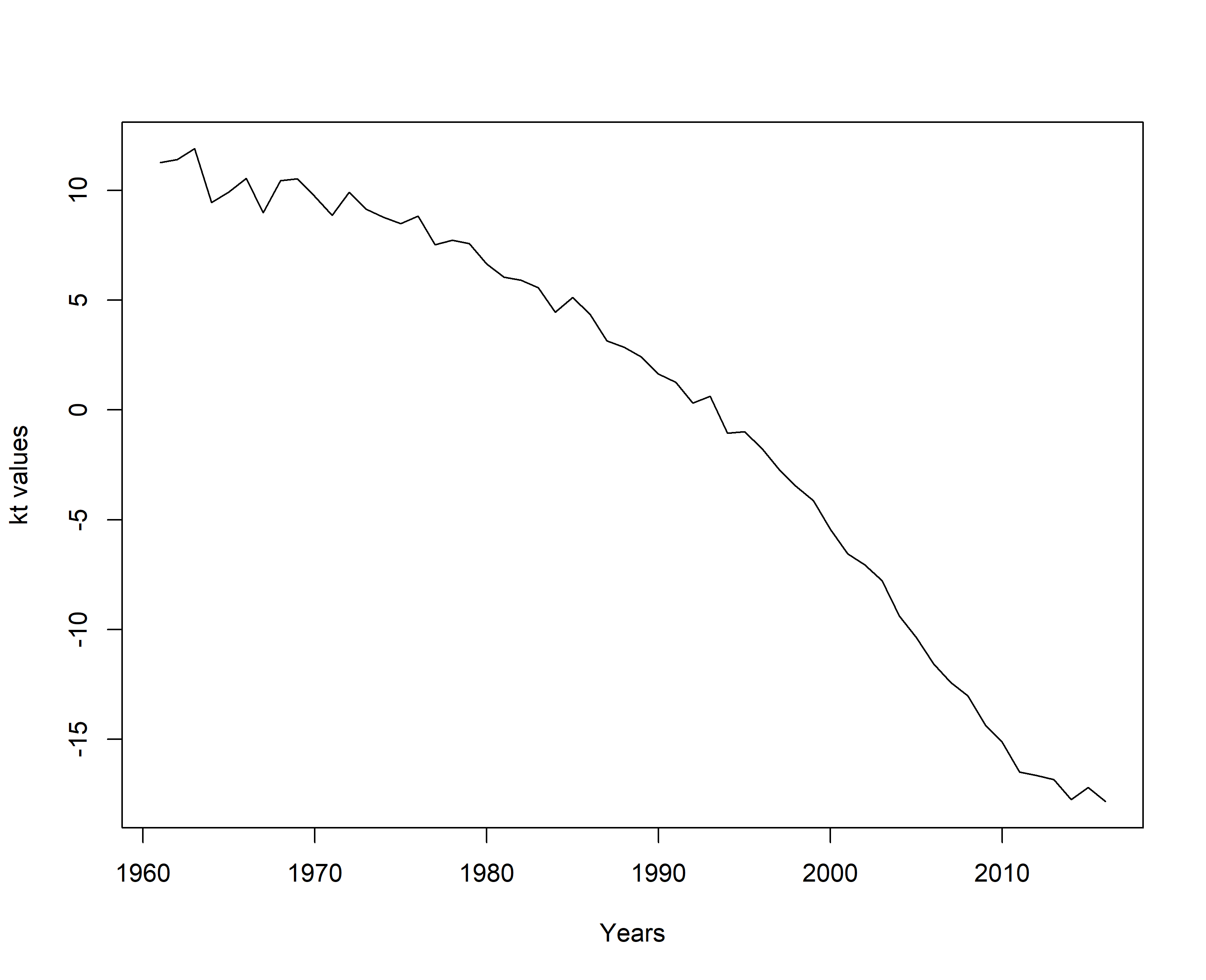}
	
	\caption{Estimated Values of $k_t^R$.}
	\label{fig1}	
\end{figure}

Then, we can write the log-likelihood of the model as follows:
$$L(D;\mu, \sigma,\eta,\alpha,\beta)=\sum_{i=1}^{T}\ln (f(r_i)).$$
The estimated $a_x^R$, $b_x^R$, $\mu, \sigma,\eta,\alpha,\beta$ parameter values are shown in Table \ref{Tab2}, while time-varying index $k_t^R$ is illustrated in Figure \ref{fig1}.
\begin{table}[h]
	\caption{Estimated Parameters for the UK}
	\centering
{\footnotesize 	\begin{tabular}{ccccccccccc}
		\toprule
		Age && $a_x$ && $b_x$ && Age && $a_x$ && $b_x$\\
		\midrule
		60 && -4.2486 && 0.0388&& 75 && -2.7879 && 0.0356\\
		61 && -4.1505 && 0.0391 && 76 &&-2.6909 &&0.0349\\
		62&& -4.0451 && 0.0399 && 77 && -2.6061 &&0.0335\\
		63&& -3.9482 && 0.0402&& 78 && -2.5122 && 0.0325\\
		64 && -3.8408 && 0.0408 && 79 && -2.4167 && 0.0314\\
		65 && -3.7472 && 0.0409 && 80 && -2.3246&& 0.0298\\
		66 && -3.6598 && 0.0401 && 81 && -2.2401 && 0.0278\\
		67 && -3.5517 && 0.0410 && 82 && -2.1366 && 0.0272\\
		68&& -3.4593&& 0.0404 && 83 && -2.0461 && 0.0257\\
		69&& -3.3607&& 0.0401 && 84 && -1.9495 && 0.0250\\
		70&& -3.2684 && 0.0392 && 85 && -1.8587 && 0.0233\\
		71&& -3.1758&&0.0378 && 86 && -1.7637 &&0.0227\\
		72&& -3.0687 && 0.0381 && 87 &&-1.6793 && 0.0213\\
		73&& -2.9749&& 0.0379 && 88 && -1.5959 && 0.0195\\
		74&& -2.8755&& 0.0369 && 89 &&-1. 5088 && 0.0179\\ 
		\midrule
		\\
		&& && &&\text{Jump Diffusion Parameters} && &&\\
		\midrule
		&&$\mu=$-0.2640 && $\sigma=$ 0.2764 && $\eta=$1.4792 &&	$\alpha=$0.0015 &&  $\beta=$0.6173\\ 		
		\bottomrule
		\label{Tab2}
	\end{tabular}}
\end{table}

Given the estimated parameters, the closed-form expression for the expected future central death rates can be derived as follows:
\begin{equation}\label{eq9}
E[\hat{m}_{xt}^R]=exp(a_x^R+b_x^R(k_0^R+(\mu-\frac{1}{2}\sigma^2)t +\sigma W(t)+ \sum_{i=1}^{N(t)}Y_i)).
\end{equation}

\subsection{Modelling the Book Population}

With the reference population in hand, it is now time to investigate the book population's mortality dynamics. Estimating the reference population first allows us to make knowledgeable decisions regarding the model's book part, and  we can also incorporate features from the reference population \cite{villegas2017}. 

The dynamics of the book population's mortality are specified through the log mortality differences of the book population and the reference population.  In this paper,  we compare the most commonly used models which are the Lee--Carter model, the age-period-cohort (APC) model, the Cairns--Blake--Dowd (CBD) model, and common age effect models to model the book population.

Note that for all the models being compared we assume that $D_{x,t}^B \sim Poisson (E_{xt}^B,q_{xt}^B)$.

\subsubsection{The Lee--Carter Model} 
The dynamics of the book population are given as follows:
\begin{equation}\label{eq10}
log(m_{x,t}^B)-log(m_{x,t}^R)=a_x^B+b_x^Bk_t^B.
\end{equation}
The term $a_x^B$ denotes the difference in the book population's level of mortality compared to that of the reference population for age $x$. Thus, we can conclude that the mortality level in the book equals $a_x^R+a_x^B$. 
Time index $k_t^B$ contributes to establishing the difference that exists in the mortality trends. The $b_x^B$ term shows us how  differences in mortality for age $x$ will respond if any change occurs in time index $k_t^B$ \cite{haberman2014}.

\subsubsection{The Common Age Effect Model}
This model may be seen as an extension of the Lee--Carter model that possesses a common age effect. It can be given by the following equation:
\begin{equation}\label{eq11}
log(m_{x,t}^B)-log(m_{x,t}^R)=a_x^B+b_x^Rk_t^B.
\end{equation}
The $a_x^B$ and $k_t^B$ parameters here are the same as in the LC model for the book population. Different from the LC model, there is a common age effect parameter, $b_x^R$, which is the same as for the reference model.

\subsubsection{The APC Model}
The APC model was introduced by Currie \cite{currie} and it is widely used in the literature. It can be regarded as a generalization of the LC model and a two-population version of this model may be written in the following way: 
\begin{equation}\label{eq12}
log(m_{x,t}^B)-log(m_{x,t}^R)=a_x^B+k_t^B+\gamma_{t-x}^B.
\end{equation}
$a_x^B, k_t^B$, and $\gamma_{t-x}^B$ respectively represent the age, the period, and the cohort effects of the book population \cite{currie}. The $\gamma_{t-x}^B$ term is utilized here in order to account for differences that exist in the cohort effect in the two populations of interest. These parameters reflect the mortality differences between the two populations.

\subsubsection{The CBD Model} Cairns et al. \cite{cairns2006} introduced the following model with the aim of fitting the mortality rates:

\begin{equation}\label{eq13}
logit(q_{x,t}^B)-logit(q_{x,t}^R)=\kappa_t^{1,B}+(x-\bar{x})\kappa_t^{2,B}.
\end{equation}

$\kappa_t^{1,B}$ and $\kappa_t^{2,B}$ are two stochastic processes and represent the time indices of book population. These parameters reflect the mortality differences between the two populations as in the APC model.

\begin{figure}[h]
	
	\centering
	\includegraphics[width=12cm, scale=1]{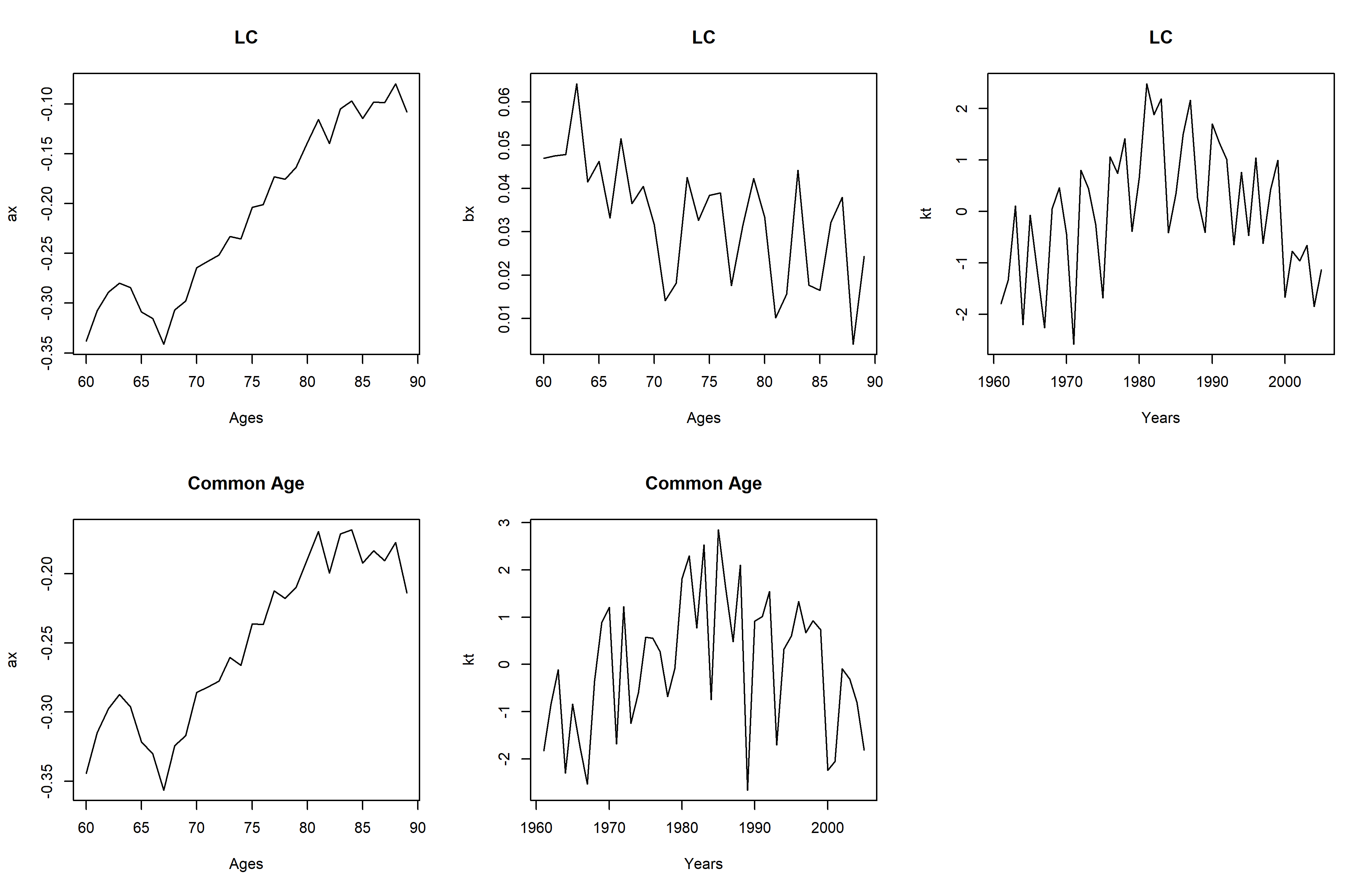}
	\includegraphics[width=12cm, scale=1]{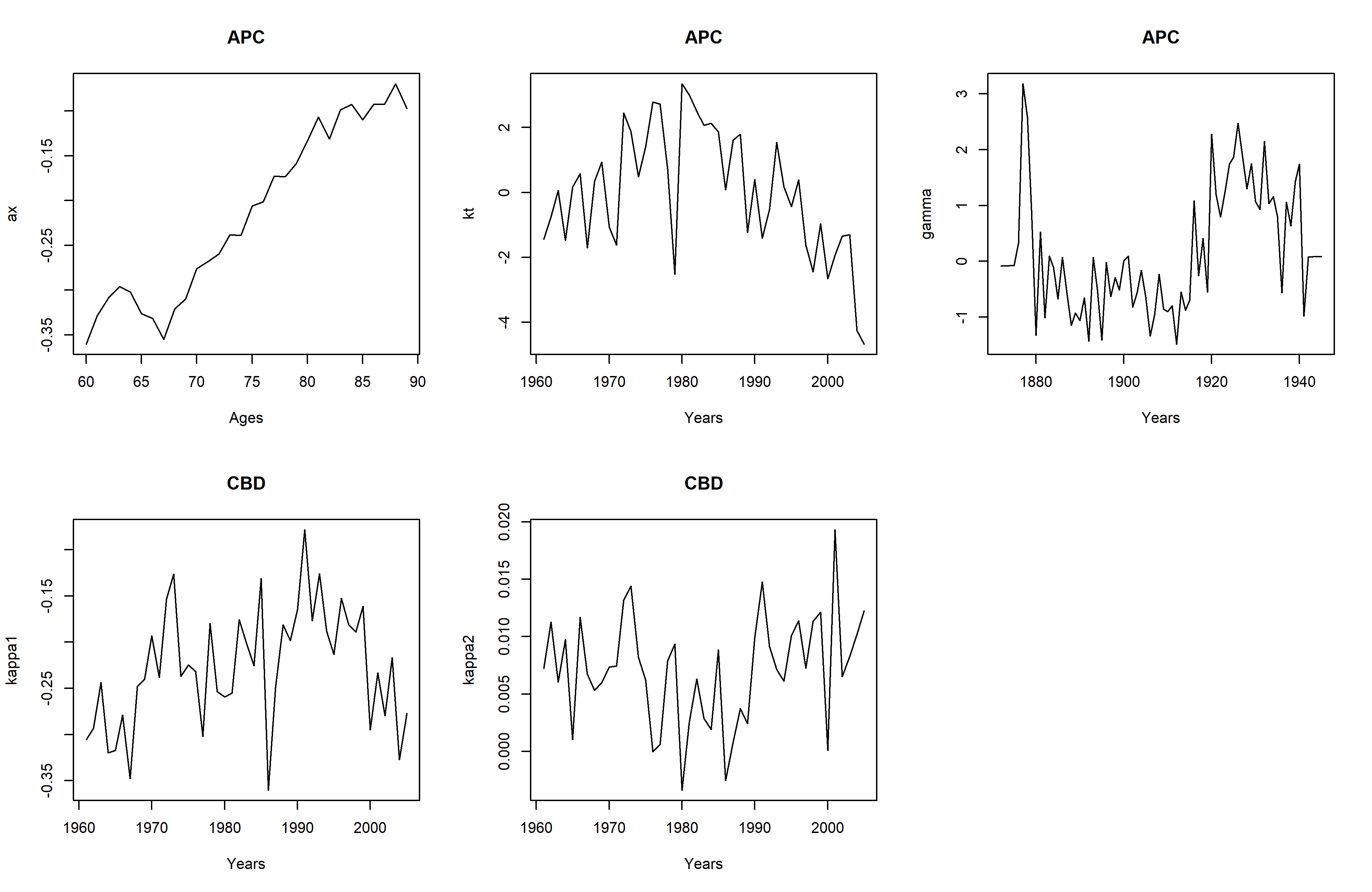}
	\caption{Estimated Parameters of the Book Population Models}
	\label{fig2}
\end{figure}

The analysis of the models considered in this section becomes something of a challenge due to the CBD model directly modelling one-year death rate $q_{x,t}$ while the others that are being considered in the paper model central death rates $m_{x,t}$. In order to compare the models in a consistent way, we must introduce an additional step for the modelling of $q_{x,t}$. We transform the one-year death probabilities in the central death rates using the identity $m_{x,t}=-\log(1-q_{x,t})$. For all mentioned models, the parameters are estimated by two main steps. As indicated before, the parameters of the book population are estimated conditional on the parameters of the reference population. Under Poisson assumption, the log-likelihood function of the book population is as follows:
$$l^B=\sum_{x,t}\big(D_{x,t}^B \ln E_{x,t}^B+D_{x,t}^B \ln m_{x,t}^B-E_{x,t}^Bm_{x,t}^B-\ln (D_{x,t}^B!)\big).$$
We estimate the parameters by applying the maximum likelihood method. The parameters obtained for the book population are given in Figure \ref{fig2} for different mortality models. 

According to Figure \ref{fig2}, the $a_x^B$ parameter shows that the younger ages reveal lower rates of mortality while the older ages reveal higher mortality. The positive values of $b_x^B$ demonstrate that mortality decreases for all ages. These results are valid for all $a_x^B$ and $b_x^B$ parameters for all mortality models of the book population. The mortality index, $k_t^B$, reflects the changes in mortality rates over the years for the LC, Common age and APC models. The $\gamma_{t-x}^B$ parameter represents the cohort related effects in the book population. The negative values of $\kappa_t^{1,B}$ parameter in the CBD model indicate the lower mortality rates while the positive values reflect the higher mortality rates. The $\kappa_t^{2,B}$ parameter controls these lower and higher mortality rates in the CBD model for the book population.

The BIC values obtained from the fitted models for book population mortality are given in Table \ref{Tab3}. The common age effect model has the lowest BIC value according to Table \ref{Tab3}. Therefore, we model the book population's mortality using the common age effect model.

\begin{table}[h]
	\caption{BIC Values for the Book Population Models}
	\centering
{\small 	\begin{tabular}{@{}cccccccccccc@{}}
		\toprule
		LC Model & \text{Common Age Effect Model} & APC Model & CBD Model \\
		\midrule
		12684.89 & \textbf{12531.63} & 12809.69 & 13759.64 \\
		\bottomrule
		\label{Tab3}
	\end{tabular}}
\end{table}

Finally, we complete the modelling framework by specifying the period's dynamics and the cohort terms, which will be used to forecast and simulate the future rates of mortality. A detailed analysis regarding the selection of the time series to be used in the dynamics can be found in the work of Li et al. \cite{lietal2015}. This part of the study confines itself to focusing on the models that are commonly applied in the literature. We assume that the two populations will experience similar improvements in the long run and thus we assume that the spread in both time indices and cohort effects should be modelled as a stationary process.

In this paper, the time-varying mortality indices of the book population $k_t^B$ are modelled as an autoregressive process of order one; we are thus able to write $k_t^B=\psi_0+\psi_1 k_{t-1}^B+\xi_t$ for the LC, the common age effect, and the APC models. In the long term, the mean of $k_t^B$ equals $\psi_0/(1-\psi_0)$ if $|\psi_1|<1$. The autocorrelation depends on the size of $\psi_1$. More technical aspects of time-series modelling can be found in the work of Tsay \cite{tsay2002}.

\subsection{Future Simulations} 

In evaluating the uncertainty of future outcomes and finding the optimal model to assess longevity basis risk, it is necessary to address all of the \textit{parameter errors}, \textit{process errors}, and \textit{model errors} from a modelling or a regulatory perspective such as that of Solvency II \cite{li2018}. \textit{Parameter error} refers to the uncertainty in estimating model parameters, while \textit{process error} arises from variations that exist within the time series. Finally \textit{model error} reflects the uncertainty that is present in the model selection.

In the literature, a number of studies have been proposed to allow for both process error and parameter error in index-based hedging. For instance, Brouhns et al. \cite{brouhns2002} used a parametric Monte Carlo simulation method for the generation of examples of model parameters following a multivariate normal distribution. Later, in a subsequent work, Brouhns et al. \cite{brouhns2005} also explored a semi-parametric bootstrapping procedure designed for the simulation of death rates from the Poisson distribution with the obtained mean equaling the observed numbers of deaths. On the other hand, Renshaw and Haberman \cite{renshaw2008} utilized fitted numbers of deaths by using the Poisson process. In another study, Koissi et al. \cite{koissi2006} used a bootstrap method for the residuals of a fitted Lee--Carter model.

Different from the existing methods, Czado et al. \cite{czado2005} and Kogure et al. \cite{kogure2009} suggested the application of Bayesian adaptations of the LC model. Li \cite{li2014} quantitatively compared possible methods for simulations; according to the conclusions of that study, sampling results will all be relatively close to each other regardless of whether the method applied is parametric, semi-parametric, Bayesian, or residual bootstrapping. All of these various simulation methods possess individual advantages and disadvantages. In this study, the bootstrapping method of Brouhns et al. \cite{brouhns2005} has been selected due to its ability to helpfully include both parameter errors and process errors in simulating future mortality rates. The bootstrapping procedure is detailed as follows:

\begin{itemize}
	\item Estimation of the parameters of the LC model is performed by using original data. Once they are obtained, those estimated parameters are then applied for estimating the numbers of deaths for both the reference and the book population by $\hat{m}_{x,t}^RE_{x,t}^R$, $\hat{m}_{x,t}^BE_{x,t}^B$.
	\item The new data on numbers of deaths are simulated from a binomial distribution for the book population to include the sampling risk and Poisson distribution is used for the reference population. The newly simulated data will then be used for estimation of the reference and book populations' mortality parameters. Incorporating this step means that the model can allow for parameter error.
	\item Next, we must fit time-series processes to the new data sample's temporal model parameters, $k_t^R$ and $k_t^B$, since we want to be able to simulate their future values. Furthermore, the inclusion of this step means that the model can allow for process error. $k_t^R$ is modelled by using the proposed model and $k_t^B$ is modelled by using AR(1).
	\item  We generate future mortality rate samples for all $x$ and future $t$ with the incorporation of the parameters obtained in step (2) and the simulated values that we gained in step (3) into $log(m_{x,t}^R)$ and $log(m_{x,t}^B)$. As a result, our set of future mortality rates will form one random future scenario.
	\item [5.] We repeat steps (1) to (4) until we have produced a total of 10,000 random future scenarios.
\end{itemize}

Different from Haberman et al. \cite{haberman2014}, in this paper, the parameter errors of the reference population has been considered by applying bootstrapping to both reference and book population models estimations.

\begin{figure}[h]	
	\centering
	\includegraphics[width=12cm, scale=1]{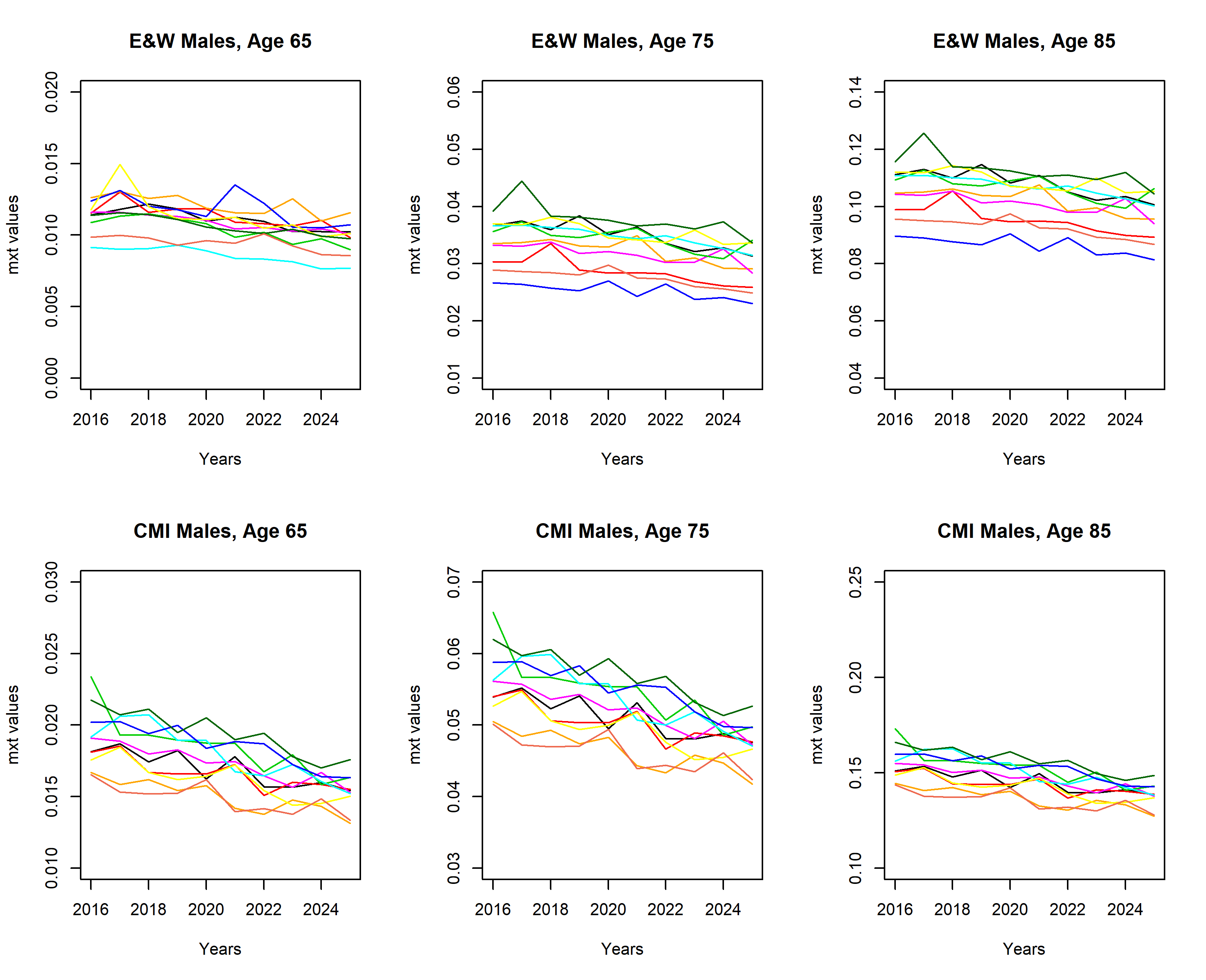}
	\caption{Sample Paths of $m_{x,t}$.}
	\label{fig3}
\end{figure}

A sample from the simulated mortality paths are shown in Figure \ref{fig3}. The mortality paths enable us to obtain projected mortality rates, hence future liabilities of pension plan and hedging instrument.

\subsubsection{Sampling Risk}

The finite sizes of the book and reference populations and the randomness of the outcomes of the individual lives cause the sampling risk. If the size of populations are infinite, the future outcomes will converge the true expected values according to the law of large numbers. Nevertheless, the size of the populations is limited in reality. Although the bigger countries have very large population sizes, the annuity or pension portfolio's size is usually small. Hence the book and reference populations' outcomes will deviate randomly from their true expected values and also from each other. To reflect the effect of the portfolio size, the number of lives is simulated as:

\begin{equation}\label{eq14}
l_{x+1,t+1}^B\sim \text{Binomial}(l_{x,t}^B,1-q_{x,t}^B)
\end{equation}

$l_{x,t}^B$ is the future number of lives aged $x$ at time $t$ of the book population. $q_{x,t}^B$ is the future mortality rate at age $x$ at time $t$ and it is simulated from the semi-parametric bootstrapping method. Simulating the number of lives of the book population by using the binomial distribution provides protection from the sampling risk \cite{li2018}.

\subsection{Comparison with the Other Mortality Models}\label{sec4.5}

After constructing a two population mortality model, we need to compare the proposed model with other mortality models and show its effectiveness. Therefore, we consider two additional two-population mortality models that are commonly used in the literature. First model is the \textit{LC model with jumps} and the second one is the \textit{LC with common age effect} model called as \textit{LC+Cohorts}. \textit{LC with jumps} model is very similar to Zhou et al. \cite{zhou2013}, however we use the relative approach to estimate the parameters of the models. Thus our mortality data could be based on different sizes of periods for reference and book population. We use the same notation with Zhou et al. \cite{zhou2013} for the \textit{LC model with jumps} and the model is as follows:
\begin{equation}\label{eq15}
\ln (m_{x,t}^i)=a_x^i+b_x^ik_t^i+e_{x,t}^i, \quad i=1,2.
\end{equation}
where $a_x^i$ and $b_x^i$ are the same as in the original LC model. The index $k_t^i$ is decomposed into the sum of two components, $k_t^i+N_t^iY_t^i$. The first component, $k_t^i$, is the time-$t$ value of an unobserved period effect index that is free of jumps, while the second term, $N_t^iY_t^i$ indicates the jump effect at time $t$. The model allows the two populations to have different jump times, jump frequencies and jump severities. They allow a maximum of one jump in a given year and the jump severity $Y_t^i$ follows a normal distribution with mean $\mu_Y$ and variance $V_Y$ (see for more model details Zhou et al. \cite{zhou2013}). 

The second mortality model that we consider here is the \textit{LC+Cohorts} model that is given in Section \ref{sec3}. The parameters of the models are estimated by using maximum likelihood method.  

The estimated parameters for these two models are given in Appendix A.

\section{Assessing Basis Risk: An Example}\label{sec5}

In this section, we consider a hedging strategy to assess longevity basis risk and to measure effectiveness of the hedge while taking mortality jumps and sampling risk into account. The effectiveness of a hedge can be described as how much longevity risk is transferred away. Following formula can be used to define the level of longevity risk reduction for the hedge as in Coughlan et al. \cite{coughlan2011}:

\begin{equation}\label{eq15}
\text{longevity risk reduction}=\bigg(1-\frac{\text{risk\ (hedged)}}{\text{risk\ (unhedged)}}\bigg)\times 100\%
\end{equation}

where the terms \textit{risk (unhedged)} and \textit{risk (hedged)} represent the appropriate dispersion-based risk measures for the aggregate longevity of the portfolio before and after the hedging. A perfect hedge would have a longevity risk reduction equal to 1 and a good hedge will have a risk reduction degree close to 1; a risk reduction degree close to 0 indicates an ineffective hedge \cite{dowdetal2019}. In this paper, the variance risk measure is used to minimise the variations in the expected future cash flows of the hedging instrument.

A simple hypothetical case study based on a pension plan is considered to illustrate the effect of the proposed mortality models and different volumes of book population data on hedge effectiveness. The pension plan members are assumed to have underlying mortality rates that are same as the CMI male assured lives dataset. Suppose all the pensioners in the plan are aged 65 and pays \pounds 1 per year on survival from ages 66 to 90. Our objective is to minimise the longevity risk exposure of the pension plan and hence we construct a hedge by using 10-year index-based longevity swap. We assume that the EW male population constitutes the floating leg's reference population, while the payments of the fixed leg of the swap are based on the CMI assured male lives. Let the interest rate be 3\% p.a. during the whole period. The current date is taken as the start of the calendar year 2016.

We use the same notation as in the work Li et al. \cite{li2018} for the hedged and the unhedged positions. The present value of the pension plan's future liability (unhedged position), $L(t)$, is given as below: 
\begin{equation}\label{eq16}
L(t)=\sum_{t=1}^{10}l_{65+t,t}^B (1+r)^{-t}
\end{equation}

As a floating-leg receiver, the present value of the longevity swap's future cash inflows, S(t), can be written as

\begin{equation}\label{eq17}
S(t)=\sum_{t=1}^{10}\big(_tp_{65}^R-_tp_{65}^{R;forward}\big)(1+r)^{-t}
\end{equation}

For this equation, we calculate random future survivor index $_tp_{65}^R$ and forward survivor index $_tp_{65}^{R;forward}$ by applying the survival probability formula, as follows: $_tp_{65}^R=(1-q_{65,0}^R)(1-q_{66,1}^R)...(1-q_{65+t-1,t-1}^R)$. Furthermore, the present value of the aggregate pension plan position after longevity hedging (hedged position) may be expressed with the following statement: 
\begin{equation}\label{eq18}
\sum_{t=1}^{10}l_{65+t,t}^B (1+r)^{-t}-w\sum_{t=1}^{10}\big(_tp_{65}^R-_tp_{65}^{R;forward}\big)(1+r)^{-t}
\end{equation}

where weight $w$ denotes the notional amount of longevity swap necessary for successful hedging to be performed \cite{li2018}. 

Moreover, in order to take the sampling risk into account, we use the binomial death process for the book population as given in Equation \eqref{eq14}. To emphasise the role of the size of the population on hedge effectiveness, we produce three simulated distributions as $l(65)=5,000$, $l(65)=10,000$ and $l(65)=100,000$. We obtain cashflows for hedged and unhedged positions for three mortality models, namely the proposed Lee-Carter model with renewal process and exponential jumps, LC model with jumps and LC+Cohorts, by considering the sampling risk. Then, the hedge effectiveness of these models are calculated.

In Table \ref{Tab4}, we present the longevity risk reduction levels when sampling risk is taken into account for the three mortality models. The results indicate that our proposed mortality jump model with renewal process provides a better risk reduction compared to the other two models. The risk reduction level increases as the sample size increases for all models which indicates that the sampling risk might be important. However, even for the smaller populations, our proposed model provides 45.07\% risk reduction while the LC model with jumps and LC+Cohorts models provide 23.17\% and 13.35\% respectively. Therefore, the analysis shows that, by using the proposed mortality model, a significant portion of the risk would be eliminated for the pension plan that exposed to mortality jump risk.

\begin{table}[h]
	\caption{Risk Reduction for Different Mortality Models}
	\centering
	\begin{tabular}{@{}cccccccccccc@{}}
		\toprule
		$l(65)$ & LC with Renewal P. \& & LC with Jumps & LC+Cohorts\\
		&   Exponential Jumps & &\\
		\midrule
		5000 & 0.4507 & 0.2317& 0.1335\\
		10000 & 0.7602 & 0.5713& 0.2605 \\
		100000 & 0.8569 &0.7392 &0.6328\\
		\bottomrule
		\label{Tab4}
	\end{tabular}
\end{table}

\section{Conclusions}\label{sec6}
Index-based hedging solutions have many advantages. In such capital market solutions, it is possible to transfer the longevity risk to capital markets at lower costs. However, the potential differences between hedging instruments and pension or annuity portfolio cause longevity basis risk. In this paper, we construct a two-population mortality model to measure and manage the longevity basis risk. 

An appropriate two-population model was built for EW male lives and CMI assured male lives to measure longevity basis risk, and the relative approach to model the populations has been adopted. The modelling process of the reference population was followed by the modelling of the dynamics of the book population's mortality. The reference population is modelled by using the LC model with renewal process and exponential jumps proposed by Özen and Şahin \cite{ozensahin} and the common age effect model outperformed among the others to model the book population. 

The bootstrap approach of Brouhns et al. \cite{brouhns2005} was applied in order to include both parameter error and process error in the simulation of future mortality rates. The Poisson distribution is used for the simulation of the reference population's lives and the binomial distribution is used for the simulation of the book population's lives to consider the sampling risk.

Furthermore, the impact of the proposed mortality model and sampling risk to hedge effectiveness is examined. For this purpose, a hypothetical pension plan and hedging strategy which consists of 10-year longevity swap is considered based on the three different two-population mortality models namely the proposed LC model with renewal process and exponential jumps, LC with jumps model and LC with common age effect model. Then the hedge effectiveness is calculated by using these three mortality models to compare the risk reduction caused by the models. The analysis suggests that the proposed mortality model provides a more effective risk reduction for mortality jump risk and sampling risk than the other two models.

A possible future study can be to construct an optimal hedging framework with collateralisation to obtain reasonable risk reduction rates by using the proposed two-population model.

\appendix

\section*{Appendix A}

In this section, the parameters of the LC model with jumps and LC+Cohorts are presented. 

The $a_x^R$ and $b_x^R$ parameters are the same for all models.

In \cite{zhou2013}, the period effect indices are modeled by following set of equations:
$$\hat{k}_{t+1}^1=\hat{k}_{t}^1+\mu_k+Z_k(t+1),$$
$$k_{t+1}^1=\hat{k}_{t+1}^1+N_{t+1}^1Y_{t+1}^1,$$
$$\hat{\Delta}_k(t)=\hat{k}_t^1-\hat{k}_t^2,$$
$$\hat{\Delta}_k(t+1)=\mu_{\Delta_k}+\phi_{\Delta_k}\hat{\Delta}_k(t)+Z_{\Delta_k}(t+1),$$
$$k_{t+1}^2=\hat{k}_{t+1}^2+N_{t+1}^1Y_{t+1}^1$$

The estimated $a_x^B$ and $b_x^B$ parameters are given in Table \ref{TabA1}.

\begin{table}[h]
	\caption{Estimated Parameters for the LC with Jumps Model}
	\centering
	\begin{tabular}{ccccccccccc}
		\toprule
		Age && $a_x$ && $b_x$ && Age && $a_x$ && $b_x$\\
		\midrule
		60 && -0.8348 && 0.0234&& 75 && -0.6154 && 0.0188\\
		61 && -0.8006 && 0.0230 && 76 &&-0.6021 &&0.0166\\
		62&& -0.7823&& 0.0217 && 77 && -0.5548&&0.0167\\
		63&& -0.7775 && 0.0222&& 78 && -0.5528 && 0.0159\\
		64 && -0.7879 && 0.0225 && 79 && -0.5282 && 0.0146\\
		65 && -0.8082 && 0.0234 && 80 && -0.4969&& 0.0153\\
		66 && -0.7920 && 0.0205 && 81 && -0.4566 && 0.0136\\
		67 && -0.8199 && 0.0199 && 82 && -0.4905 && 0.0126\\
		68&& -0.7798&& 0.0199 && 83 && -0.4426&& 0.0131\\
		69&& -0.7650&& 0.0194 && 84 && -0.4414 && 0.0124\\
		70&& -0.7193 && 0.0189 && 85 && -0.4493 && 0.0124\\
		71&& -0.6876&&0.0177 && 86 && -0.4449 &&0.0140\\
		72&& -0.6941 && 0.0191 && 87 &&-0.4244 && 0.0131\\
		73&& -0.6655&& 0.0175 && 88 && -0.3931 && 0.0106\\
		74&& -0.6572&& 0.0168 && 89 &&-0.4136 && 0.0113\\ 		
		\bottomrule
		\label{TabA1}
	\end{tabular}
\end{table}

The parameters of the jump component of the model are presented in Table \ref{TabA2}.
\begin{table}[h]
	\caption{Estimated Parameters for the LC with jumps model}
	\centering
	\begin{tabular}{ccccccccccc}
		\toprule
		$\mu_k=-0.4973$ &$\mu_Y^1=4.2915$ & $\mu_Y^2=4.5614$&$\mu_{\Delta_k}=-0.3108$\\[0.5ex]
		$\phi_{\Delta_k}=0.0496$ &$V_Y^1=0.5608$&$V_Y^2=0.6849$& $V_Z=0.3915$\\

		\bottomrule
		\label{TabA2}
	\end{tabular}
\end{table}

The probabilities of jump frequencies are $Pr(N_t^1=0,N_t^2=0)=0.7763$, $Pr(N_t^1=0,N_t^2=1)=0.0967$ and $Pr(N_t^1=1,N_t^2=1)=0.1269$.

\begin{figure}[h]
	
	\centering
	\includegraphics[width=10cm, scale=1]{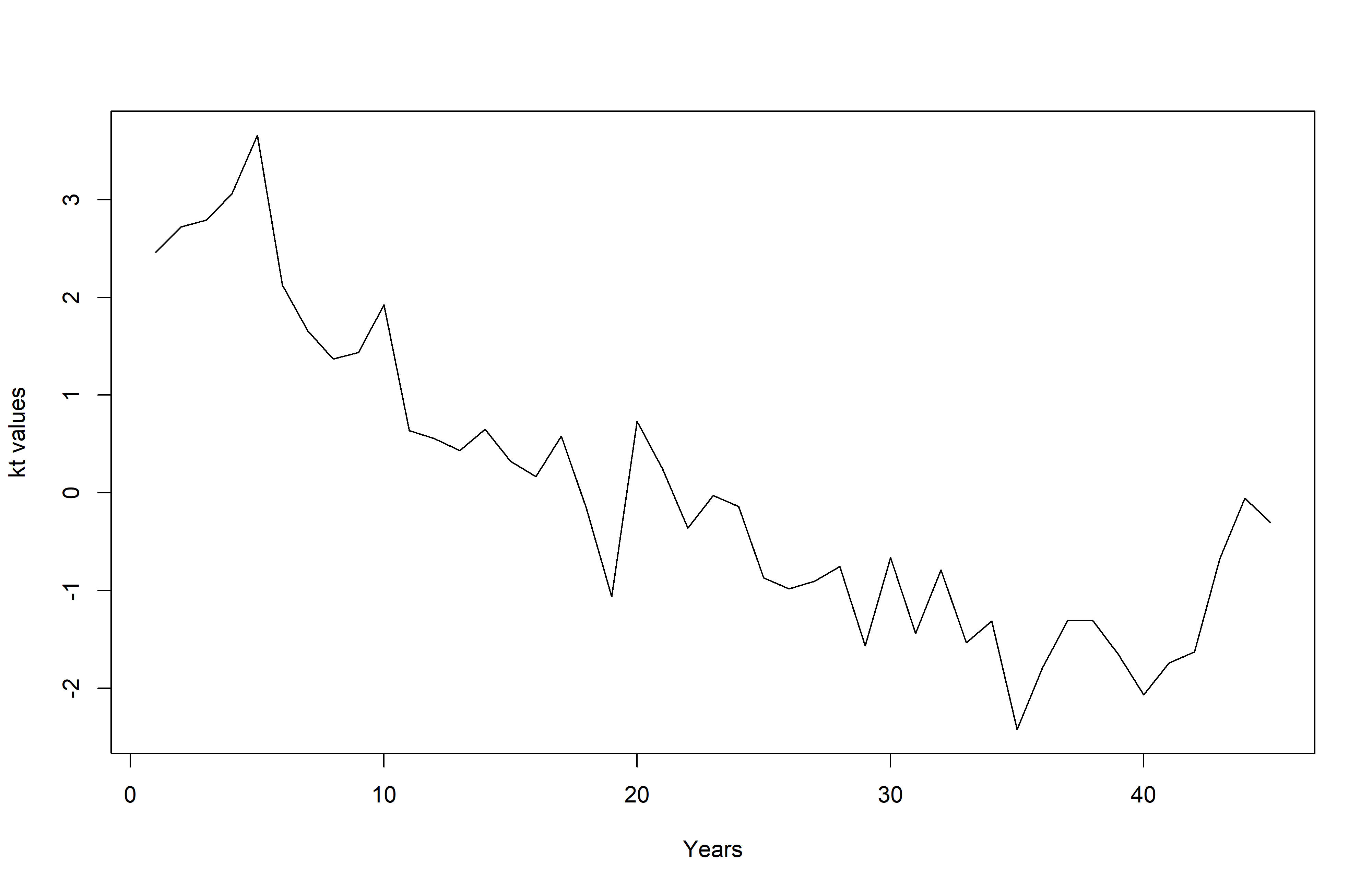}
	\caption{Estimated $k_t$ Values of Book Population}
	\label{figA1}
\end{figure}

The parameters of the LC+Cohorts model book population are presented in Table \ref{A3} and Figure \ref{figA1}.
\begin{table}[h]
	\caption{Estimated Parameters for the Book Population of LC+Cohorts Model}
	\centering
	\begin{tabular}{ccccccccccc}
		\toprule
		Age && $\hat{a_x}$  && Age && $\hat{a_x}$ \\
		\midrule
		60 && -0.5431 && 75 && -0.3930  \\
		61 && -0.5123 && 76 &&-0.3886  \\
		62&& -0.4981&&  77 && -0.3545 \\
		63&& -0.4897 && 78 && -0.3569  \\
		64 && -0.4995  && 79 && -0.3419  \\
		65 && -0.5207  && 80 && -0.3171\\
		66 && -0.5223   && 81 && -0.2893  \\
		67 && -0.5495  && 82 && -0.3201 \\
		68&& -0.5135  && 83 && -0.2828 \\
		69&& -0.5032  && 84 && -0.2801  \\
		70&& -0.4664  && 85 && -0.2988  \\
		71&& -0.4513 && 86 && -0.2904\\
		72&& -0.4500  && 87 &&-0.2846  \\
		73&& -0.4293  && 88 && -0.2639 \\
		74&& -0.4287  && 89 &&-0.2944 \\ 
		
		\bottomrule
		\label{A3}
	\end{tabular}
\end{table}
\end{document}